\documentstyle[12pt]{article}
\newcommand{\address}[1]{\centerline{\small\it #1}}
\newcommand{\draft}{}
\renewcommand{\title}[1]{\centerline{\Large\bf #1}\par\bigskip}
\renewcommand{\author}[1]{\centerline{\large #1}}
\renewcommand{\maketitle}{}
\newcommand{\pacs}[1]{}
\newcommand{\narrowtext}{}
\newcommand{\widetext}{}
\begin{document}
\draft
\title{Classical XY Model in 1.99 Dimensions\footnote{
Published in Phys. Rev. Lett. {\bf 74}, 3916 (1995).
}}
\author{Tohru Koma$^*$ and Hal Tasaki$^\dagger$}
\address{Department of Physics, Gakushuin University,
Mejiro, Toshima-ku, Tokyo 171, JAPAN}
\maketitle
%%%%%%%%%%%%%%%%%%%%%%%%%%%%%%%%%%%%%%%%%%%%
\begin{abstract}
We consider the classical XY model ($O(2)$ nonlinear $\sigma$-model)
on a class of
lattices with the (fractal) dimensions $1<D<2$.
The Berezinskii's
harmonic approximation suggests that the model undergoes a phase
transition in which the low temperature phase is characterized by
stretched exponential decay of correlations.
We prove an exponentially decaying upper bound
for the two-point correlation functions at non-zero temperatures,
thus excluding the possibility of such a phase transition.
\end{abstract}
%%%%%%%%%%%%%%%%%%%%%%%%%%%%%%%%%%%%%%%%%%%
\narrowtext
The classical XY model
(or, equivalently, $O(2)$ nonlinear $\sigma$-model)
has been a subject of considerable interest in the contexts
of statistical physics and relativistic field theory.
The model exhibits a standard ferromagnetic phase transition
in dimensions $d>2$,
while it undergoes an exotic phase transition called
the Berezinskii-Kosterlitz-Thouless transition
\cite{Berezinskii,KT,FS}
in $d=2$.
The existence of a phase transition in two dimensions
is in a remarkable contrast with the $O(n)$-rotator models ($\sigma$-models)
with $n\ge3$ such as the classical Heisenberg model,
which are conventionally
believed to be asymptotically free
and have no phase transitions in two dimensions \cite{n>2}.
In one dimension, general arguments
guarantee that the  XY model
(like all the other short range spin systems) has no phase transitions.

It is now common to
regard the dimension $d$ as a continuous parameter,
having in mind lattices with fractal structures for example.
Then a natural question is whether the
classical XY model
exhibits a (finite temperature)
phase transition in the intermediate dimensions $1<d<2$.
Although there have been no publications directly devoted to
this problem
as far as we know, we believe that the problem is important and
is worth settling.

To illustrate that the problem is nontrivial, denote by $T_{\rm c}(d)$
the critical temperature
(at which the susceptibility diverges)
of the classical XY model on the $d$-dimension hypercubic lattice.
The rigorously established facts are that $T_{\rm c}(d)$ is finite for
$d=2,3,4,\cdots$, and is vanishing for $d=1$.
If one naively regards $T_{\rm c}(d)$ as a function of the continuous parameter
$d$, then the most natural ``interpolation'' may be
that  $T_{\rm c}(d)$ continuously decreases
below $d=2$, and vanishes continuously at $d=1$ (or some dimension $1<d<2$).
A much stronger support for this naive guess comes from the observation that
a straightforward extension of the Berezinskii's harmonic
approximation \cite{Berezinskii} (which correctly predicts the
existence of the Kosterlitz-Thouless transition in $d=2$)
indicates that, in $1<d<2$, there can be an exotic low temperature
phase characterized by stretched exponential decay of correlations.
One who came across with this argument might
conjecture the existence of a phase transition in $1<d<2$.

In the present Letter, we address the above problem
and provide a conclusive result.
We prove that, in the classical XY model
defined on a
lattice with the (fractal) dimension $1\le D<2$,
the two-point correlation function decays exponentially at any
finite temperature.
Our (fractal) dimension $D$ coincides with the euclidean dimensions
for regular lattices.
This result excludes the possibility of the
new type of phase transition mentioned above \cite{Chirality}.
It also leads us to a somehow unexpected conclusion that $T_{\rm c}(D)$, as
a function
of continuous dimension $D$, drops discontinuously
\cite{Thouless} to zero at $D=2$.

The proof is carried out in two steps.
First we use the McBryan-Spencer method \cite{MS}
to prove an upper bound for the two-point correlation function
in terms of a stretched exponentially decaying function.
This result automatically extends to
a very general class of models with
continuous symmetries, and strengthens the well-known
result that these systems have no long range order in $d\le2$
\cite{MW,Tasaki,frac2D}.
Next we apply the Simon's inequality and his argument
on the decay of correlation \cite{Simon} to conclude
that the correlation decays exponentially in the classical XY model.

%%%%%%%%%%%%%%%%%%%%%%%%%%%%%%%%%%%%%%%%%%%%%%%%%%%%%%%%%
\paragraph*{Definitions:}
We consider a general connected lattice
$\Lambda=(\Lambda_{\rm s},\Lambda_{\rm b})$,
where $\Lambda_{\rm s}$ is a set of countably infinite sites
$x,y,\ldots$,
and $\Lambda_{\rm b}$ is a set of bonds, i.e.,
pairs of sites $(x,y),(u,v),\ldots$ through which spin variables directly
interact.
We define the classical XY model on $\Lambda$.
The Hamiltonian (action) is given by
\begin{equation}
{\cal H}
= - \sum_{(u,v)\in \Lambda_{\rm b}}
{\bf s}_u \cdot {\bf s}_u
= - \sum_{(u,v)\in \Lambda_{\rm b}}
\cos(\theta_u-\theta_v),
\label{action}
\end{equation}
where ${\bf s}_x=(\cos\theta_x,\sin\theta_x)$
is the spin variable at site
$x \in \Lambda_{\rm s}$.
The two-point correlation function
at inverse temperature $\beta\ge0$ is
\begin{equation}
{\langle {\bf s}_x\cdot {\bf s}_y \rangle}_\beta
:=
{Z_\beta}^{-1}
\left(\prod_z \int_{-\pi}^\pi d\theta_z\right)
e^{-\beta{\cal H}}
\cos(\theta_x-\theta_y)
\label{correlation}
\end{equation}
with the partition function
\begin{equation}
Z_\beta
:=
\left(\prod_z \int_{-\pi}^\pi d\theta_z\right)
e^{-\beta{\cal H}}.
\label{Z}
\end{equation}
The expectation value (\ref{correlation}) is to be interpreted as
the thermodynamic limit of the corresponding finite volume quantities.

%%%%%%%%%%%%%%%%%%%%%%%%%%%%%%%%%%%%%%%%%%%%%%%%%%%%%%%%%
\paragraph*{Berezinskii's argument:}
Before describing our results, we shall briefly review
the Berezinskii's argument and discuss its extension.
The essence of the Berezinskii's harmonic approximation
\cite{Berezinskii,MS,Tsallis} is to
make the model into a Gaussian model by
replacing $\cos(\theta_u-\theta_v)$ in the Hamiltonian (\ref{action})
by $[1-(\theta_u-\theta_v)^2/2]$ and replacing the range of integrals
in (\ref{correlation}) and (\ref{Z}) to $(-\infty,\infty)$.
A naive expectation is that the approximation gives sensible results
for large $\beta$ where local difference of fields
$(\theta_u-\theta_v)$ is small \cite{BereNote}.
After making the above replacements
(and suitably eliminating contributions of a zero-mode),
the correlation function is evaluated as
\begin{eqnarray}
 &&
 {\langle {\bf s}_x\cdot {\bf s}_y \rangle}_\beta^{\rm Gauss}
 =
 \frac{\left(\prod_z \int_{-\infty}^\infty d\theta_z\right)
 e^{-\beta\widetilde{\cal H}}
 \cos(\theta_x-\theta_y)}
 {\left(\prod_z \int_{-\infty}^\infty d\theta_z\right)
 e^{-\beta\widetilde{\cal H}}}
 \nonumber\\
 &&
 = \exp[-\langle(\theta_x-\theta_y)^2\rangle_\beta^{\rm Gauss}/2]
 = \exp[\beta^{-1}C(x,y)/2],
 \label{harmonicAp}
\end{eqnarray}
where
$\widetilde{\cal H}=
\sum_{(u,v)\in\Lambda_{\rm b}}(\theta_u-\theta_v)^2/2
= -\sum_{u\in\Lambda_{\rm s}}\theta_u(\Delta\theta)_u/2$
is the approximate Gaussian Hamiltonian.
Here the lattice Laplacian is defined as
$(\Delta f)_z = \sum_{y\in\Lambda_{\rm s}:(y,z)\in\Lambda_{\rm b}}
(f_y-f_z)$.
The quantity
$C(x,y)=-\beta\langle(\theta_x-\theta_y)^2\rangle_\beta^{\rm Gauss}$
can be characterized as follows.
For fixed $x$, $y$, let $\varphi_z$ be the solution of the lattice Poisson
equation
\begin{equation}
 -(\Delta\varphi)_z= \delta_{x,z}-\delta_{y,z}.
 \label{Poisson}
\end{equation}
Then $C(x,y)$ is equal to the ``potential difference''
$C(x,y)=\varphi_y-\varphi_x$.
By examining the solution of (\ref{Poisson})
in a general $d$-dimensional lattice, we find that $C(x,y)$
behaves for large $|x-y|$ as
\begin{equation}
 C(x,y) \simeq\left\{
 \begin{array}{ll}
 A(d)|x-y|^{2-d}-B(d) & \mbox{if $2<d$}\\
 -A(d)\log|x-y| & \mbox{if $d=2$}\\
 -A(d)|x-y|^{2-d} & \mbox{if $1\le d<2$}
 \end{array}
 \right.,
 \label{Psol}
\end{equation}
where $A(d)$ and $B(d)$ are strictly positive constants which depend
only on the lattice structure. The asymptotic behavior (\ref{Psol})
for nonintegral $d$ was obtained by extrapolating from those for
integral $d$.
On a fractal lattice, one should interpret $d$ in (\ref{Psol})
as a suitable ``fractal dimension'' \cite{specNote}.

The approximation (\ref{harmonicAp})
along with  (\ref{Psol})
implies that the asymptotic behavior as $|x-y|\uparrow\infty$
of the correlation functions for large $\beta$ (low temperatures)
is given by
${\langle {\bf s}_x\cdot {\bf s}_y \rangle}_\beta
\simeq \sigma(\beta)^2>0$ for $d>2$,
${\langle {\bf s}_x\cdot {\bf s}_y \rangle}_\beta
\approx |x-y|^{-\eta(\beta)}$ for $d=2$,
and
${\langle {\bf s}_x\cdot {\bf s}_y \rangle}_\beta
\approx \exp[-|x-y|/\xi(\beta)]$ for $d=1$.
Here (the approximations for) the order parameter $\sigma(\beta)$,
the critical exponent $\eta(\beta)$,
and the correlation length $\xi(\beta)$
are given by
$\sigma(\beta)=\exp[-B(d)/(4\beta)]$,
$\eta(\beta)=A(2)/(2\beta)$,
and $\xi(\beta)=2\beta/A(1)$.
These estimates for correlations
(including the $\beta$-dependence of $\sigma(\beta)$, $\eta(\beta)$,
and $\xi(\beta)$)
recover the known (or expected) behavior in a semi-qualitative
manner.
It is remarkable that the simple approximation yields
such strong results.

In dimensions $1<d<2$, the same approximation
yields the asymptotic behavior
\begin{equation}
 {\langle {\bf s}_x\cdot {\bf s}_y \rangle}_\beta
 \approx
 \exp[-\alpha(\beta)|x-y|^{2-d}],
\label{stretchedasym}
\end{equation}
with a positive function $\alpha(\beta)$ of $0\le\beta<\infty$,
which is usually referred to as a stretched exponential decay.
Since the correlation decays exponentially for sufficiently
small $\beta$, this observation suggests that
the classical XY model undergoes a phase transition
in $1<d<2$.

It is interesting that, as we will show in Lemma,
the McBryan-Spencer argument leads us rigorously to
an upper bound for the correlation
decaying by a similar stretched exponential law.

%%%%%%%%%%%%%%%%%%%%%%%%%%%%%%%%%%%%%%%%%%%%%%%%%%%%%%%%%
\paragraph*{Main results:}
We define the ``sphere'' $S_n(x)$ of radius $n$
centered at $x\in \Lambda_{\rm s}$
by
\begin{equation}
S_n(x):=\{y\in \Lambda_{\rm s} | {\rm dist}(x,y)=n\}.
\end{equation}
Here ${\rm dist}(x,y)$ is the graph-theoretic distance
between the sites $x$, $y$, which is
defined as the minimum number of bonds in $\Lambda_{\rm b}$ that
one needs to connect $x$ and $y$.
We assume that there exists a ``(fractal) dimension" $D$
of the lattice \cite{D} such that the number of sites in $S_n(x)$ is
bounded uniformly from above as
\begin{equation}
\mathop{\rm sup}_{x\in\Lambda_{\rm s}} |S_n(x)|\le Cn^{D-1}
\label{assumeD}
\end{equation}
with a positive constant $C$.
It is obvious that the dimension $D$ coincides with the
euclidean dimension if $\Lambda$ is a regular lattice.

Then our main result is the following exponentially decaying upper bound
for the two-point correlation.

{\em Theorem}---If the dimension $D$ satisfies $1\le D<2$,
the two-point correlation function is bounded as
\begin{equation}
\left|\langle {\bf s}_x \cdot {\bf s}_y \rangle_\beta\right|
\le \exp[-m(\beta) \{{\rm dist}(x,y)-R(\beta)\}]
\end{equation}
for any $0\le\beta<\infty$,
where $m(\beta)>0$ and $R(\beta)>0$
are functions of $\beta$ \cite{long}.

In the proof of the theorem, we make use of the following lemma.
The lemma states an upper bound for the two-point correlation
in terms of a stretched exponentially decaying function
like the one obtained from the Berezinskii's approximation.

We note that the following lemma can be trivially extended to cover
more general class of systems with a global continuous
symmetry.
The examples include $O(n)$ ($n>2$) rotators (nonlinear $\sigma$-model)
\cite{MS}, quantum Heisenberg models \cite{Itoh},
and the Hubbard model \cite{KomaTasaki} for itinerant electrons.
The main theorem, on the other hand, can be proved only
in the classical XY model for the technical reason that the Simon
inequality is known only for one- and two-component spin systems.
We suspect that the statement of the theorem is valid for
the larger class of models.

{\em Lemma}---For $1\le D<2$, we have
\begin{equation}
\left|\langle {\bf s}_x \cdot {\bf s}_y \rangle_\beta\right|
\le \exp\left[-f(\beta)({\rm dist}(x,y)^{2-D}-1)\right]
\label{stexpdecay}
\end{equation}
for any $0\le\beta<\infty$,
where the function $f(\beta)$ is given by
\begin{equation}
 f(\beta)=
 \max_{q\ge 0}\left[q-\beta C^2(\cosh q-1)/(2-D)\right]
 >0,
 \label{f}
\end{equation}
and is decreasing in $\beta$.

%%%%%%%%%%%%%%%%%%%%%%%%%%%%%%%%%%%%%%%%%%%%%%%%%%%%%%%%%
\paragraph*{Proof of Lemma:}
We fix two sites $x$, $y$ throughout the proof.
Following McBryan and Spencer \cite{MS},
we make the complex transformations
\begin{equation}
\theta_z \rightarrow \theta_z + i \phi_z
\end{equation}
in the numerator of (\ref{correlation}), where $\phi_z$ are real
numbers to be determined.
This means that we deform
the path of integration and use the periodicity of the cosine
to cancel contributions of the lateral contours.
The transformation combined with $|e^{iz}|=1$ for real $z$ yields
%%%%%%%%%%%%%%%%%%%%%%%
\widetext
\begin{eqnarray}
&&
\left|\langle {\bf s}_x \cdot {\bf s}_y \rangle_\beta\right|
\le
e^{-(\phi_x-\phi_y)} Z_\beta^{-1}
\left(\prod_z \int_{-\pi}^\pi d\theta_z\right)
\exp\!\!\left[\beta \sum_{(u,v)\in \Lambda_{\rm b}}
\cos(\theta_u-\theta_v)\cosh(\phi_u-\phi_v)\right]
\nonumber \\
&&
\le
e^{-(\phi_x-\phi_y)}
\exp\!\!\left[\beta \sum_{(u,v)\in \Lambda_{\rm b}}
\left(\cosh(\phi_u-\phi_v)-1\right)\right] .
\label{MSbound}
\end{eqnarray}
\narrowtext
%%%%%%%%%%%%%%%%%%%%%%%
Following the idea of Picco \cite{Picco},
we choose $\{\phi_z\}$ as
\begin{equation}
\phi_z =
\left\{
\begin{array}{ll}
    q(N^{2-D}-1)       &  \mbox{if $z=x$}          \\
    q(N^{2-D}-{\rm dist}(x,z)^{2-D}) &  \mbox{if $1\le{\rm dist}(x,z)\le N$}\\
    0                  &  \mbox{if ${\rm dist}(x,z) > N $}
\end{array}
\right.
\label{choice}
\end{equation}
where $N={\rm dist}(x,y)$, and $q$ is a positive
constant to be determined. From the choice (\ref{choice}),
we have that
\begin{equation}
\phi_x-\phi_z=0
\label{n=1}
\end{equation}
for $z$ such that ${\rm dist}(x,z)=1$, and that
\begin{equation}
|\phi_u-\phi_v|\le q n^{1-D}\le q
\label{nnbound}
\end{equation}
for $u,v$ satisfying $n={\rm dist}(x,u)={\rm dist}(x,v)-1\ne 0$.

The summation in the exponential in the right-hand side of
(\ref{MSbound}) can be evaluated as
%%%%%%%%%%%%%%%%%%%%%%%
\widetext
\begin{eqnarray}
&&
\sum_{(u,v)\in \Lambda_{\rm b}}
\left(\cosh(\phi_u-\phi_v)-1\right)
\nonumber\\
&&
= \sum_{n=1}^{N-1} \sum_{u:{\rm dist}(x,u)=n}
\mathop{\sum_{v:(u,v)\in \Lambda_{\rm b}}}_{
{\rm dist}(x,v)\ge {\rm dist}(x,u)}
\left[\cosh(\phi_u-\phi_v)-1\right] \nonumber \\
&&
\le \sum_{n=1}^{N-1} \sum_{u:{\rm dist}(x,u)=n}
\mathop{\sum_{v:(u,v)\in \Lambda_{\rm b}}}_{
{\rm dist}(x,v)\ge {\rm dist}(x,u)}
(\cosh q-1)q^{-2}(\phi_u-\phi_v)^2 \nonumber \\
&&
\le C(\cosh q-1) \sum_{n=1}^{N-1} \sum_{u:{\rm dist}(x,u)=n}
n^{2-2D}
\nonumber \\
 &&
\le C^2(\cosh q-1)\sum_{n=1}^{N-1} n^{D-1}\times n^{2-2D}
\nonumber \\
&&
\le C^2(\cosh q-1)(N^{2-D}-1)/(2-D) ,
\label{bigbound}
\end{eqnarray}
\narrowtext
%%%%%%%%%%%%%%%%%%%%%%%
where we have used (\ref{n=1}) and the choice (\ref{choice})
of $\{\phi_z\}$ to rewrite the summation at the first line.
The succeeding three bounds have been obtained by using
(\ref{nnbound}) and (\ref{assumeD}).
By substituting (\ref{bigbound}) and (\ref{choice}) into
the main bound (\ref{MSbound}),
we get a bound of the form (\ref{stexpdecay}).
Finally we optimize the bound by choosing the constant $q$
according to (\ref{f}).

%%%%%%%%%%%%%%%%%%%%%%%%%%%%%%%%%%%%%%%%%%%%%%%%%%%%%%%%%
\paragraph*{Proof of Theorem:}
We assume $1<D<2$.
Recall the Simon inequality \cite{Simon}
for the classical XY model
\begin{equation}
\langle {\bf s}_x \cdot {\bf s}_y \rangle_\beta
\le \beta \mathop{\sum_{u \in V, v \notin V}}_{(u,v)\in
\Lambda_{\rm b}}
\langle {\bf s}_x \cdot {\bf s}_u \rangle_\beta
\langle {\bf s}_v \cdot {\bf s}_y \rangle_\beta ,
\label{Simon}
\end{equation}
where the finite set $V$ is chosen so that $x\in V$, and $y\notin V$.
We set $V=S_R(x)$ with some $R<{\rm dist}(x,y)$.
By using
(\ref{assumeD}) and the bound (\ref{stexpdecay}) of the lemma,
(\ref{Simon}) implies
\begin{equation}
\langle {\bf s}_x \cdot {\bf s}_y \rangle_\beta
\le e^{-m(\beta) R}
\langle {\bf s}_{v_1} \cdot {\bf s}_y \rangle_\beta
\label{bound1}
\end{equation}
with
\begin{equation}
e^{-m(\beta) R}=
C^2 \beta R^{D-1} \exp\left[-f(\beta)(R^{2-D}-1)\right] ,
\label{defmass}
\end{equation}
where the site $v_1$ gives the maximum value of
$\langle {\bf s}_v \cdot {\bf s}_y \rangle_\beta$ in the summation in
the right-hand side of (\ref{Simon}).
Noting that the right-hand side of (\ref{defmass}) is less than $1$ for
sufficiently large $R$, we let $R(\beta)$ be the value of $R$ which
maximizes the ``mass" $m(\beta)$ in the left-hand side of
(\ref{defmass}).
We note that, for such $R(\beta)$ and
for any given distance $r={\rm dist}(x,y)$,
there exist non-negative integers
$\ell$ and $r^\prime$ satisfying $r=R(\beta) \ell +r^\prime$
and $r^\prime<R(\beta)$.
(For $r<R(\beta)$, we set $r'=r$ and $\ell=0$.)
By setting $R=R(\beta)$,
we can apply the Simon inequality (\ref{Simon}) to the right-hand side of
(\ref{bound1}) repeatedly at least $(\ell-1)$ more times.
We thus obtain
the desired exponentially decaying bound \cite{noteTasaki} as
\begin{equation}
\langle {\bf s}_x \cdot {\bf s}_y \rangle_\beta
\le e^{-m_\beta R(\beta)\ell}
\langle {\bf s}_{v_\ell} \cdot {\bf s}_y \rangle_\beta
\le e^{-m_\beta (r-r^\prime)} .
\label{bound2}
\end{equation}

\par\bigskip
{\bf Note added\footnote{
Phys. Rev. Lett. {\bf 75}, 984 (1995).
}:}
After the publication of the paper, we learned that the XY model on a class
of finitely
ramified fractals is studied in \cite{Dhar}, which is one of the pioneering
papers in
``physics on fractals.''
It is a pleasure to thank Deepak Dhar and Tetsuya Hattori for useful
correspondences.

%%%%%%%%%%%%%%%%%%%%%%%%%%%%%%%%%%%%%%%%%%%%%%%%%%%%%%%%%


\begin{thebibliography}{99}
\bibitem{Chirality}
Our result does not exclude
the possibility of phase transitions which do not manifest
itself in decay properties of the spin-spin correlation
function.
An example is the chirality transition
which breaks the
$Z_2\cong O(2)/SO(2)$ symmetry.
\bibitem{BereNote}
The approximation is not well controlled in the sense that
one has to make these two replacements together to get a sensible
result.
See \cite{MS} for a critical review of the approximation.
\bibitem{specNote}
Although the relevant fractal dimension is the one determined by
the properties of the lattice Laplacian (i.e., a free random walker),
it is not the spectral dimension.
The dimension in (\ref{Psol}) depends on the choice of metric $|x-y|$
while the spectral dimension is independent of metrics.
\bibitem{D}
The dimension $D$ defined here is distinct from that
appeared in (\ref{Psol}).
Nevertheless we believe that the inequality $D<2$ with the present definition
provides the optimal criterion for the absence of phase transition.
We also note that (\ref{assumeD}) may not hold in some random fractal
lattices, where one cannot find the constant $C$ which is independent
of $x$.
\bibitem{long}
It is possible to extend the theorem to the models with  the Hamiltonian
${\cal H}=-\sum_{u,v\in\Lambda}J_{u,v}\,{\bf s}_u\cdot{\bf s}_v$
with the interaction $J_{u,v}$ satisfying
$J_{u,v}\ge0$ and
$\sup_u\sum_v J_{u,v} \{{\rm dist}(u,v)\}^2 <\infty$.
This means that moderately long ranged interactions do not
change the behavior of the model for $D<2$.
\bibitem{noteTasaki}
An application of the Simon inequality to fractal spin systems
was discussed in \cite{Tasaki}.
Rather involved argument in \cite{Tasaki} can be simplified by
following the logic presented here.
\end{thebibliography}
\end{document}